\documentclass{aa}  

\usepackage{graphicx}
\usepackage{txfonts}
\begin{document}

   \title{The impact of Lyman-$\alpha$ radiative transfer on large-scale clustering in the Illustris simulation}


   \author{C. Behrens
          \inst{1,2}
          \and
          C. Byrohl
          \inst{2,3}
          \and
          S. Saito
          \inst{3}
          \and
          J. C. Niemeyer
          \inst{2}          
          }

   \institute{Scuola Normale Superiore, Piazza dei Cavalieri 7, I-56126 Pisa, Italy\\
              \email{christoph.behrens@sns.it}
         \and
             Institut fuer Astrophysik, Georg-August Universit\"at Goettingen,
             Friedrich-Hund-Platz 1, 37075 Goettingen, Germany\\
        \and
             Max-Planck-Institut f\"{u}r Astrophysik, Karl-Schwarzschild-Star{\ss}e 1, D-85740 
             Garching bei M\"{u}nchen, Germany\\
             }

   \date{Received 16 August 2017; accepted 5 February 2018}

 
  \abstract
   {Lyman-$\alpha$ emitters (LAEs) are a promising probe of the large-scale structure at high redshift, $z\gtrsim 2$. In particular, the Hobby-Eberly Telescope Dark Energy Experiment aims at observing LAEs at 1.9 $<z<$ 3.5 to measure the baryon acoustic oscillation (BAO) scale and the redshift-space distortion (RSD). However, it has been pointed out that the complicated radiative transfer (RT) of the resonant Lyman-$\alpha$ emission line generates an anisotropic selection bias in the LAE clustering on large scales, $s\gtrsim 10\,{\rm Mpc}$. This effect could potentially induce a systematic error in the BAO and RSD measurements. Also,  there exists a recent claim to have observational evidence of the effect in the Lyman-$\alpha$ intensity map, albeit statistically insignificant.}
   {We aim at quantifying the impact of the Lyman-$\alpha$ RT on the large-scale galaxy clustering in detail. For this purpose, we study the correlations between the large-scale environment and the ratio of an apparent Lyman-$\alpha$ luminosity to an intrinsic one, which we call the `observed fraction', at $2<z<6$.}
   {We apply our Lyman-$\alpha$ RT code by post-processing the full Illustris simulations. 
   We simply assume that the intrinsic luminosity of the Lyman-$\alpha$ emission is proportional to the star formation rate (SFR) of galaxies in Illustris, yielding a sufficiently large sample of LAEs to measure the anisotropic selection bias.}
   {We find little correlation between large-scale environment and the observed fraction induced by the RT, and hence a smaller anisotropic selection bias than has previously been claimed. We argue that the anisotropy was overestimated in previous work due to insufficient spatial resolution; it is important to keep the resolution such that it resolves the high-density region down to the scale of the interstellar medium, that is, $\sim1$ physical kpc. %
   We also find that the correlation can be further enhanced by assumptions in modeling intrinsic Lyman-$\alpha$ emission.}
   {}


   \maketitle
%

\section{Introduction}
High-redshift $z\gtrsim 2$ galaxies with prominent Lyman-$\alpha$ emission \citep{Partridge1967}, referred to as Lyman-$\alpha$ emitters (LAEs), have become the subject of intense research over the last two decades. Since LAEs are believed to be powered at least partially by ongoing star formation, they are expected to belong to a relatively low-mass and young class of actively star-forming galaxies. This suggests that the number density of LAEs is large enough to map out the large-scale structure of the high-redshift universe. Increasing numbers of LAEs are being observed by surveys including narrow-band imaging \citep[e.g.,][]{Nakajima2012} and integral field unit spectroscopy such as MUSE \citep[e.g.,][]{MUSE}, amounting to about 10$^4$ emitters known to date since the first detections \citep[e.g.,][]{Cowie1998,Hu1996}. These observations enable us to study galaxy clustering at somewhat small scales \citep[e.g.,][]{Diener2017,Ouchi2010,Ouchi2017} as well as evolution of the Lyman-$\alpha$ luminosity function \citep[e.g.,][]{Konno2016,Konno2017,Ouchi2008}. A remarkable example for such surveys is the Hobby-Eberly Telescope Dark Energy Experiment \citep[hereafter, HETDEX]{Adams2011,Hill2008}. HETDEX will map out the three-dimensional distribution of nearly one million LAEs \citep{Leung2017} at $1.9 < z < 3.5$ over $400\,{\rm deg^2}$. This allows us to precisely measure the large-scale ($\gtrsim 10\,{\rm Mpc}$) clustering of LAEs \citep[][]{Agrawal2017,Chiang2013} and of the Lyman-$\alpha$ intensity map (Saito et al., in preparation) with the primary scientific goal being to determine the cosmic expansion history via baryon acoustic oscillations (BAOs) and the growth of structure via redshift-space distortions (RSD). We refer the reader to \citet{Alam2016} and references therein for recent efforts of BAO and RSD measurements. In addition, these surveys offer exciting opportunities to study the connection between LAEs and their environment, including the cross-correlation between LAEs and the Lyman-$\alpha$ forest or other galaxy populations.

However, one important challenge in using LAEs for cosmology arises due to the relatively complex nature of the radiative transfer (RT). The Lyman-$\alpha$ line is a resonant line, and hence even small column densities of neutral hydrogen can scatter Lyman-$\alpha$ photons near the line center numerous times \citep[][]{Zheng2002}. In principle, the RT effect impacts on all scales; within the intergalactic medium (IGM) or circumgalactic medium (CGM), as well as in the interstellar medium (ISM) within LAEs\footnote{We note that here we explicitly define and distinguish the terms IGM, CGM, and ISM in terms of spatial scales. IGM spreads over a relatively linear regime, $\gtrsim 1\,{\rm Mpc}$, while CGM is found in the fully nonlinear regime, on the scales of dark matter halos, i.e.,$\lesssim 1\,{\rm Mpc}$. ISM exists on scales of $\lesssim 10\,{\rm kpc}$.}. In particular, \citet[][hereafter, ZZ11]{Zheng2011} pointed out that the RT effect through intervening gas clouds, even in the IGM or CGM, could lead to an additional, non-gravitational selection bias when using LAEs as a probe of the large-scale structure \citep[also see][for details on their simulation]{Zheng2010}. Post-processing a large cosmological simulation with a Lyman-$\alpha$ RT code, ZZ11 find a significant correlation between the line-of-sight velocity gradient and the observed fraction of Lyman-$\alpha$ emission. Specifically, they find that emitters in a large-scale environment exhibiting a large line-of-sight velocity gradient are preferentially detected over emitters not living in such an environment. They measure this bias by comparing the luminosity a virtual observer would infer given by the flux obtained from an emitter (after RT) divided by the intrinsic luminosity of the source, which we refer to as \textit{`the observed fraction'} $\epsilon$ here. They find this quantity to vary with velocity gradient by about an order of magnitude, and subsequently show that this anisotropically biases the measured two-point correlation function (2PCF), effectively elongating its shape along the line of sight. This effect can, in principle, cause a systematic error in measuring the radial BAO and RSD measurement from the anisotropic galaxy clustering \citep{Greig2012,Wyithe2011a}. Intriguingly, \citet{Croft2016} claimed  observational evidence (albeit not statistically significant) of such a large-scale elongation in the cross-correlation signal between the Lyman-$\alpha$ intensity map and quasar distribution even at $z\sim 2$, while ZZ11 performed their simulation analysis at $z\sim 6$.

It is therefore important to understand how the RT couples the intrinsic properties of Lyman-$\alpha$ emitting galaxies to their cosmological environment. For this purpose, we use a suite of modern galaxy-formation simulations in a cosmological context. We post-process the publicly available Illustris simulation \citep{Vogelsberger2014,Nelson2015} to simulate the possible anisotropic selection bias in the LAE clustering due to the non-linear nature of the Lyman-$\alpha$ RT. 

In \cite{Behrens2013} (in the following: BN13), we used a high-resolution cosmological simulation from the MareNostrum collaboration \citep{Ocvirk2008} and subsequent RT post-processing to validate this claim. However, we did not find the strong correlation ZZ11 detected, and also found no significant deformation of the 2PCF. Analytic estimates by \cite{Wyithe2011a} suggest that this non-detection could be due to statistics, since the MareNostrum simulation was about one eighth of the comoving size of the simulation ZZ11 employed. In this paper, we revisit these results using the well-known Illustris suite of high-resolution simulations to investigate the relation between the large-scale environment and the observability of LAEs in a cosmological context. To directly compare with the results of ZZ11, we not only consider redshifts within the range of HETDEX, but also a high-redshift snapshot from the simulation at $z=5.85$. 

We note that Lyman-$\alpha$ radiation transport has already been applied to parts of the Illustris simulation by \citet{Gronke2017}. While they focused on extended Lyman-$\alpha$ emission from massive halos and modeled the emission from these objects in great detail, they did apply the RT to selected halos only by effectively cutting out small subvolumes around the relevant galaxies. In our work, we consider a comparatively simple scheme for determining the emission of Lyman-$\alpha$ photons, but in turn process the full simulation volume and all emitters inside.

The paper is structured as follows. In Sect. \ref{sec:methods}, we briefly describe how we converted the Illustris data into a format suitable for the post-processing RT step, and the RT code used as far as it has not been described elsewhere. In Sect. \ref{sec:results}, we present our results on the correlations between large-scale environment and the observed fraction of Lyman-$\alpha$ radiation, followed by the concluding remarks in Sect. \ref{sec:conclusions}. In the appendix, we show the robustness of our results and the LAE sample in more detail.

\section{Methods}
\label{sec:methods}
We make use of the public release of the Illustris dataset as described in \cite{Nelson2015}, a set of high-resolution hydrodynamical simulations of a large volume ($\sim$ 107 Mpc) including recipes for cooling, star formation, and feedback. The Illustris suite fits well a large number of observations, including the star formation history, stellar mass functions, and the distribution of spirals and ellipticals, among others, and has been applied to a large number of applications. For more details, we refer the reader to \cite{Vogelsberger2014,Vogelsberger2014a} and to the detailed documentation available on the website of the Illustris project\footnote{www.illustris-project.org} from which the data were obtained.

Specifically, we use the density, ionization, temperature, and velocity information from various snapshots of the Illustris-1 simulation, and the Friend-of-Friends (FoF) group catalogs for the halo/emitter positions. We note that while catalogs for subhalos exist and we have run RT simulations using their positions instead of the halo positions in the FoF catalog, we find the results to be similar as far as the scope of this paper -- correlations with the large-scale structure -- is concerned. Additionally, we make use of the star formation rates (SFRs) from the halo catalogs in the Illustris simulation to estimate the intrinsic Lyman-$\alpha$ luminosity of each central galaxy (see below).

\subsection{Preprocessing}
The Illustris simulations are based on the moving mesh scheme of the Arepo code \citep{Springel2010} whereas the Monte-Carlo (MC) radiative transfer (RT) code employed here relies on patch-based adaptive mesh refinement (AMR), that is, a hierarchy of fixed grids. A snapshot of the Illustris simulations contains a large number of dark-matter particles and hydrodynamical tracer particles. To convert these to our target AMR format, we employ an intermediate step. First, we use an octet-based tree to estimate the final refinement structure. Since our main constraint in running the simulations is the final size of the AMR dataset in memory, we choose to refine a cell of the octet tree if it contains more than 32 particles. Secondly, we distribute the particles' mass, momentum, and internal energy onto the refined octet tree using a cloud-in-cell algorithm. Finally, we convert the octet tree to a patch-based AMR structure suitable for reading by the BoxLib\footnote{ Developed by the Center for Computational Sciences and Engineering at Lawrence Berkeley National Laboratory, see https://ccse.lbl.gov/BoxLib/} \citep{Zhang2016} infrastructure. We make sure that, in this last step, every cell refined in the octet tree is also refined in the patch-based AMR structure. We choose our refinement to make sure that we reach a target resolution of $\sim$ 1 physical kpc (pkpc). We carry out these steps for a number of snapshots at different redshifts between $z=2.00$ and $z=5.85$ (details are given in Table \ref{table1}). For comparison with ZZ11, we also run some snapshots with an artificially reduced spatial resolution comparable to that of ZZ11.

\begin{table}
\caption{Redshift $z$, spatial resolution $\Delta$, and number of LAEs considered, $N_{LAE}$, for the post-processed snapshots. We consider halos with SFR $>$ 0.1 M$_\odot$/yr and M$_{h,200}$ $>$ 10$^{10}$ M$_\odot$ as LAEs. The intermediate resolution runs have a resolution similar to ZZ11 at redshift 5.85. For each redshift, we also state the average neutral fraction, $f_{\rm IGM}$, at a characteristic hydrogen number density of $10^{-4}$ cm$^{-3}$. }             
\label{table1}      
\centering          
\begin{tabular}{l l l}     
\hline\hline       
     $z$ & Resolution $\Delta$ [pkpc] & $N_{LAE}$\\ 
                   
\hline                 
2.00 ($f_{\mathrm{IGM}}=2 \times 10^{-5}$) & & \\
high resolution & 1.2 & 45594 \\  
\hline    
3.01 ($f_{\mathrm{IGM}}=3.7 \times 10^{-5}$)& & \\
high resolution & 0.8 & 45434\\
intermediate resolution  & 51.9 & 45434\\
\hline                 
4.01 ($f_{\mathrm{IGM}}=6.8 \times 10^{-5}$)& & \\
high resolution & 0.7 &39782 \\
\hline    
5.85 ($f_{\mathrm{IGM}}=35 \times 10^{-5}$)& & \\
high resolution & 0.5 & 23114 \\   
intermediate resolution & 30.4 & 23114 \\   
low resolution & 121.5 & 23114 \\
  
\hline                  
\end{tabular}
\end{table}

\subsection{Modelling Lyman-$\alpha$ emitters}
Similarly to previous works, we assume that Lyman-$\alpha$ emission is dominated by recombination after ionization by far UV radiation from young stars. Given our resolution of $\sim$ 1 kpc, we assume that these stars reside in the cores of galaxies, embedded in the innermost part of a dark matter halo. The Lyman-$\alpha$ emission is therefore placed at the center of a dark matter halo. We assume each halo of mass M$_{h,200}$ $>$ 10$^{10}$ M$_\odot$ to emit Lyman-$\alpha$ photons, with M$_{h,200}$ being the mass enclosed in a sphere with density $\rho = 200\rho_c$, and $\rho_c$ the critical density.  Additionally, we restrict ourselves to considering only halos with SFR $>$ 0.1 M$_\odot$/yr and assume an intrinsic Lyman-$\alpha$ luminosity \citep[e.g.,][]{Furlanetto2006}
\begin{equation}
L_{int} = \frac{\mathrm{SFR}}{\mathrm{M}_\odot/\mathrm{yr}} \times  10^{42} \mathrm{erg/s}
.\end{equation}
As stated above, we ignore halos below the mass and SFR threshold. The initial line profile of the emission is set to be Gaussian around the line center, with the width given by the virial temperature of the halo. The line profile is an important factor, since different choices for it can alter the effect on the optical depth seen in the IGM \citep{Laursen2011}. We decided to choose a Gaussian profile in line with previous work, and because it requires no additional parameters.

We emphasize that while we use the SFR obtained from the hydro simulations, ZZ11 and BN13 instead approximated the SFR using the halo mass \citep[see][]{Trac2007a} and used this to calculate the intrinsic Lyman-$\alpha$ luminosity
\begin{equation}
L_{int}^{ZZ} = 0 .68 \frac{\mathrm{M_{h}}}{10^{10} \mathrm{M}_\odot} \times  10^{42} \mathrm{erg/s}
.\end{equation}
Additionally, they used a lower mass cutoff, and measured the halo mass as the mass inside a sphere of 200 times the mean density of the Universe. We have checked that this different definition of the halo mass does not affect our results. Our choices for the fiduclal model are motivated by the fact that we have robust SFRs from the Illustris simulation. Furthermore, given our spatial resolution of $\sim$ 1 kpc, we want to make sure that the gas distribution and kinematics in and around galaxies we consider as LAEs are spatially resolved, thus we use a rather high cut-off in mass. When comparing the SFR-halo mass relation with that from \cite{Trac2007a}, we find that while their result implies a power-law slope of 1 for SFR($M_h$), we find a slope of 1.7-1.9 for the halos in the Illustris simulations.

The escape of Lyman-$\alpha$ photons from a highly inhomogeneous interstellar ISM remains a hard problem to solve in simulations \cite[e.g.][]{Hansen2006,Laursen2013,Gronke2016a,Gronke2016b}. These inhomogeneities are not captured sufficiently well even by high-resolution simulations, as parsec or even sub-parsec resolution would be required to resolve the ISM structure. As has been discussed by \cite{Gronke2016b} in great detail, the observed spectra of LAEs deviate from synthetic ones as a consequence. As in this paper we are primarily concerned with the question of the anisotropic selection bias due to RT, which is an effect related to the coupling between photons and the IGM, we have decided to artificially remove the effect of the ISM on the escaping photons. We render the ISM transparent to Lyman-$\alpha$ photons by removing the gas in cells with number densities above $\rho_{\mathrm{cut}}=0.13 cm^{-3}$, the threshold for star-forming gas in the Illustris simulations. We stress that this definition of the star-forming ISM does not include all of the ISM, defined earlier by its spatial scales. For rendering the star-forming ISM transparent, we were primarily motivated by the fact that it is the star-forming ISM that is mostly affected by radiative feedback and turbulence, rendering it very difficult to accurately compute radiative transfer at the available resolution.
This method tends to over-predict the nongravitational bias since processing in the ISM moves photons out of the line center, 
gradually decoupling photons from the large-scale environment by suppressing scattering in the IGM. In the appendix, we compare our results with cases in which we included the ISM. 

Additionally, and similar to previous work, we do not include dust in our simulations. Since we are interested in learning about the maximum impact of the radiative transfer on the observed large-scale distribution of LAEs, this approach is appropriate, in particular given the severe physical uncertainties in modeling the effect of dust \citep[see][and references therein]{Asano2013,Aoyama2017}.

\subsection{Lyman-$\alpha$ RT}
We use our Lyman-$\alpha$ code to solve the RT problem, based on the \textit{BoxLib} library and the \textit{Nyx} code \citep{Almgren2013}. It was used previously in \cite{Behrens2014} and \cite{Behrens2014b}. For this work, we 
additionally include the redshifting of photons due to the Hubble flow, periodic boundaries, and the peeling-off method to efficiently generate surface-brightness maps. For details, we refer the reader to BN13 and references therein. Here, we briefly summarize the simulation technique. 

We use the Monte-Carlo approach typically used for Lyman-$\alpha$ RT, probing the gas, velocity, and temperature distribution by a large number of tracer photons. These photons are launched at the center of halos and propagate in initially random directions. They penetrate an optical $\tau_D$ drawn from an exponential distribution before they interact with the gas. The optical depth along their path is integrated taking into account the local gas density, velocity, and temperature. Additionally, the redshifting of photons due to the Hubble flow is taken into account by calculating the Hubble parameter at the redshift of the snapshot $H(z)$, and increasing the wavelength of the photon by a factor $\propto H(z)d$, with $d$ being the distance to the last point of scattering (or, initially, the distance to the point of emission). Since we do not consider dust, scattering on hydrogen atoms is the only process considered. If a point of interaction is reached, the photon is scattered coherently in the frame of the scattering atom, changing both the frequency and the direction of propagation of the photon in the frame of the observer. A new $\tau_D$ is drawn and the process is repeated until the photon is considered to have left the volume; since we apply periodic boundary condition, we need to specify this leaving criteria. Similarly to previous works, we follow photons through the simulation volume only for a fraction of the box size. If a photon is more than d$_{cut}$ away from its source, it is discarded and counted as escaped. The underlying assumption is that photons are typically shifted out of resonance a few Mpc away from their source due to the Hubble flow. We check this assumption and other details of the employed RT scheme in the appendix. 

Using the so-called peeling-off scheme or next-event estimator \citep{Yusef-Zadeh1984,Zheng2002}, at each scattering event the probability that the photon is scattered in some specified direction of observation and escapes in a single flight is calculated. This probability, the position of the scattering, the frequency, and various other properties are recorded for each such event. Additionally, these quantities are also recorded for the initial spawning of each photon, which corresponds to the possibility that a photon could escape without being scattered at all. Due to the peeling-off scheme, each photon potentially contributes to the flux leaving the simulation at each scattering event (and at the initial emission).
\subsection{Analysis of the RT simulations}
In the literature, the term escape fraction is frequently, but not always, defined to be the fraction of photons escaping from a galaxy without being absorbed. With this definition, 
formally, the escape fraction in our simulations is equal to 1, because we do not include dust or other processes that can destroy Lyman-$\alpha$ photons. 

However, some photons are scattered into the line of sight of the observer at large projected distances from their emitter and will become part of a diffuse background. We therefore need to specify how to calculate the fraction of emission that is actually detected by a synthetic observer and associated with a galaxy at a given position. Again, we call this fraction \textit{the observed fraction}, $\epsilon$, throughout our paper. To calculate it, we chose to employ the same method as ZZ11/BN13; we bin the contributions of all the photons in the simulation onto a fine grid (using 4108$^2$ bins to obtain a similar\footnote{To be more precise, this choice results in obtaining the same angular resolution at redshift 5.85 as ZZ11.} resolution to the one used in ZZ11). We have checked that our results do not change if we, instead of keeping the number of pixels constant, enforce a constant angular resolution over the redshifts in range. Physically, our angular resolution is 1.0/0.8/0.7/0.6'' at redshifts 2/3.01/4.01/5.85. Given a LAE position, we use the Friends-of-Friends algorithm to find all the pixels above a certain threshold in surface brightness, beginning with the pixel in which the halo position falls. If this pixel is above the threshold, we test whether any of the four neighboring pixels are above the threshold as well; if so, we continue to test their direct neighbors, and so on. This approach has the advantage that it adapts to different intrinsic sizes and luminosities of the emitters, where, for example, a fixed-size aperture integration would ignore such differences. 

In case the surface brightness of a pixel onto which an LAE is projected is not above the threshold, we nevertheless assign the spectrum and luminosity of this pixel to the LAE. By default, we use a threshold of 1.8$\times$10$^{-19}$ erg/s/cm$^2$/arcsec$^2$. To avoid double-counting photons, we assign the luminosity of a pixel only to one halo, even if there are multiple halos projected onto the same pixel. Thus, once a pixel's luminosity is linked to a specific LAE, all following LAEs projected onto this pixel will have no luminosity. The choice of our threshold was motivated by previous work. However, we checked the influence of its exact value on our results and found that it primarily affects the mean observed fraction, but not the trends in question here.

Using this algorithm, we can calculate the total flux associated with a halo, and the total luminosity L$_\mathrm{app}$ that a synthetic observer would infer from the received flux. Together with information about the intrinsic Lyman-$\alpha$ luminosity of the source L$_\mathrm{int}$, we can calculate the observed fraction $\epsilon$,
\begin{equation}
\epsilon = \frac{L_\mathrm{app}}{L_\mathrm{int}}
.\end{equation}

\section{Results}
\label{sec:results}
\subsection{Luminosity function}
\begin{figure}
   \centering
   \includegraphics[width=\hsize]{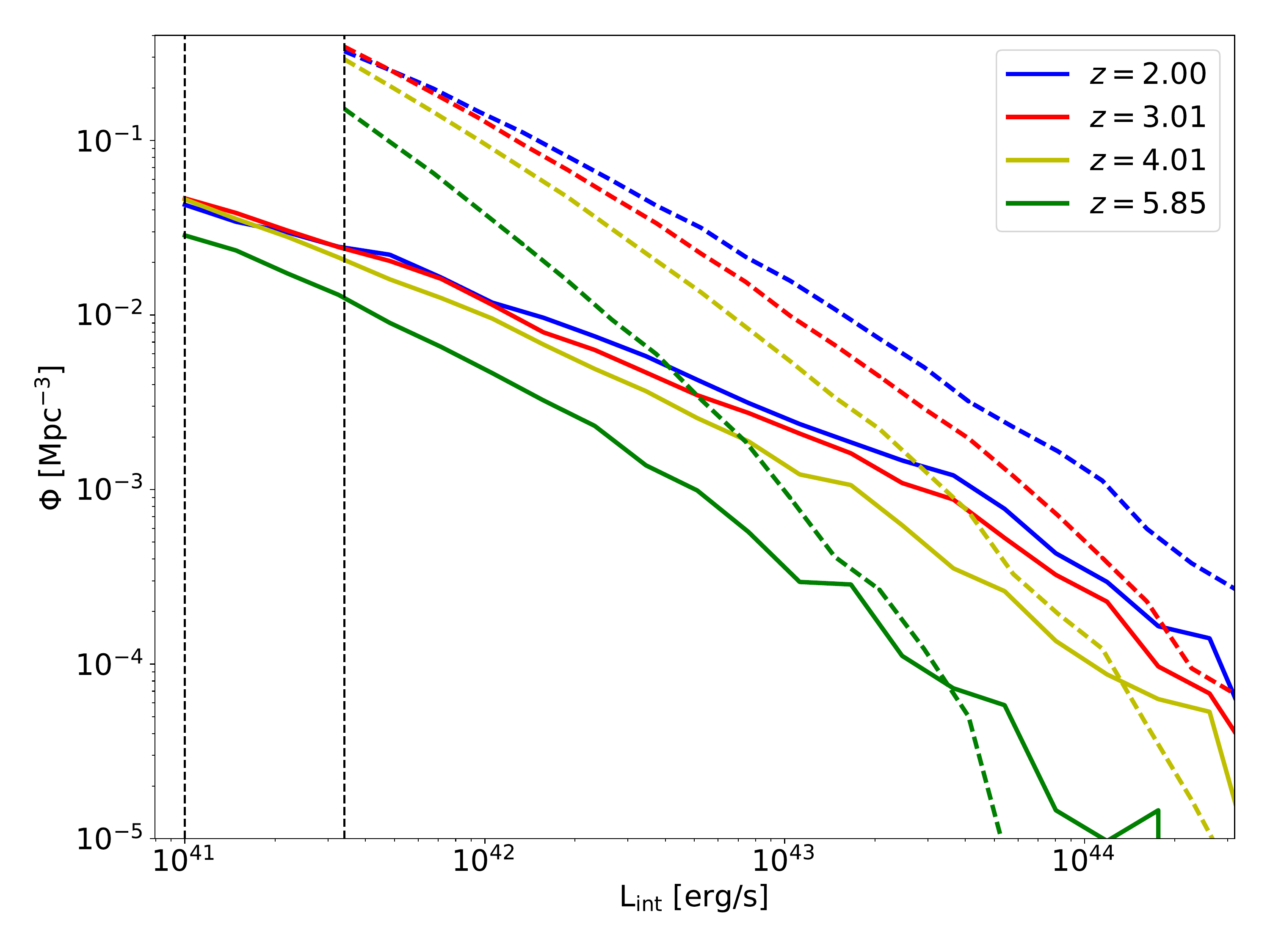}
      \caption{Intrinsic LAE luminosity function for different snapshots. Solid lines: Fiducial model utilizing the SFR data from the Illustris simulation. Dashed lines: The model used by ZZ11 and BN13, for comparison. The vertical lines mark the cutoff in the LFs expected from the SFR/mass cuts chosen.}
      \label{fig:intrinsicLF}
\end{figure}

In Fig. \ref{fig:intrinsicLF}, we show the intrinsic luminosity function (LF) of the LAEs for snapshots at different redshifts, that is, the luminosity in Lyman-$\alpha$ that we assign prior to radiative transport. Our fiducial model is shown with solid lines while the dashed lines show the emission model used by  ZZ11/BN13. The strict cutoffs at low luminosities result from the minimum star formation and minimum mass requirement, respectively, as explained above. 

As is evident from the plot, the total luminosity density is higher for the ZZ model, mainly due to higher luminosities assigned to low-mass emitters. This results in a different slope which is also reflected in the luminosity functions after RT processing (see below). As a consistency check, one can calculate the total cosmic SFR density (CSFRD) for the two models, which in the case of our fiducial choice reduces to integrating the SFRs given in the Illustris snapshot. While Illustris was calibrated to match the CSFRD relation (at least down to $z=1$) and the calculated CSFRD therefore matches the expectations (e.g., with a value of $\sim$ 10$^{-2}$ M$_\odot$/Mpc$^{-3}$ at $z=5.85$), the ZZ11/BN13 model overpredicts the CSFRD by a factor of a few, for example, by a factor of 5 at redshift 5.85. On the other hand, as will be shown below, the intrinsic LF at redshift 5.85 is already lower than the observed one, probably owing to variations in the SFR-halo mass relation. We therefore do not expect to match the observed LFs in detail.

\begin{figure}
   \centering
   \includegraphics[width=\hsize]{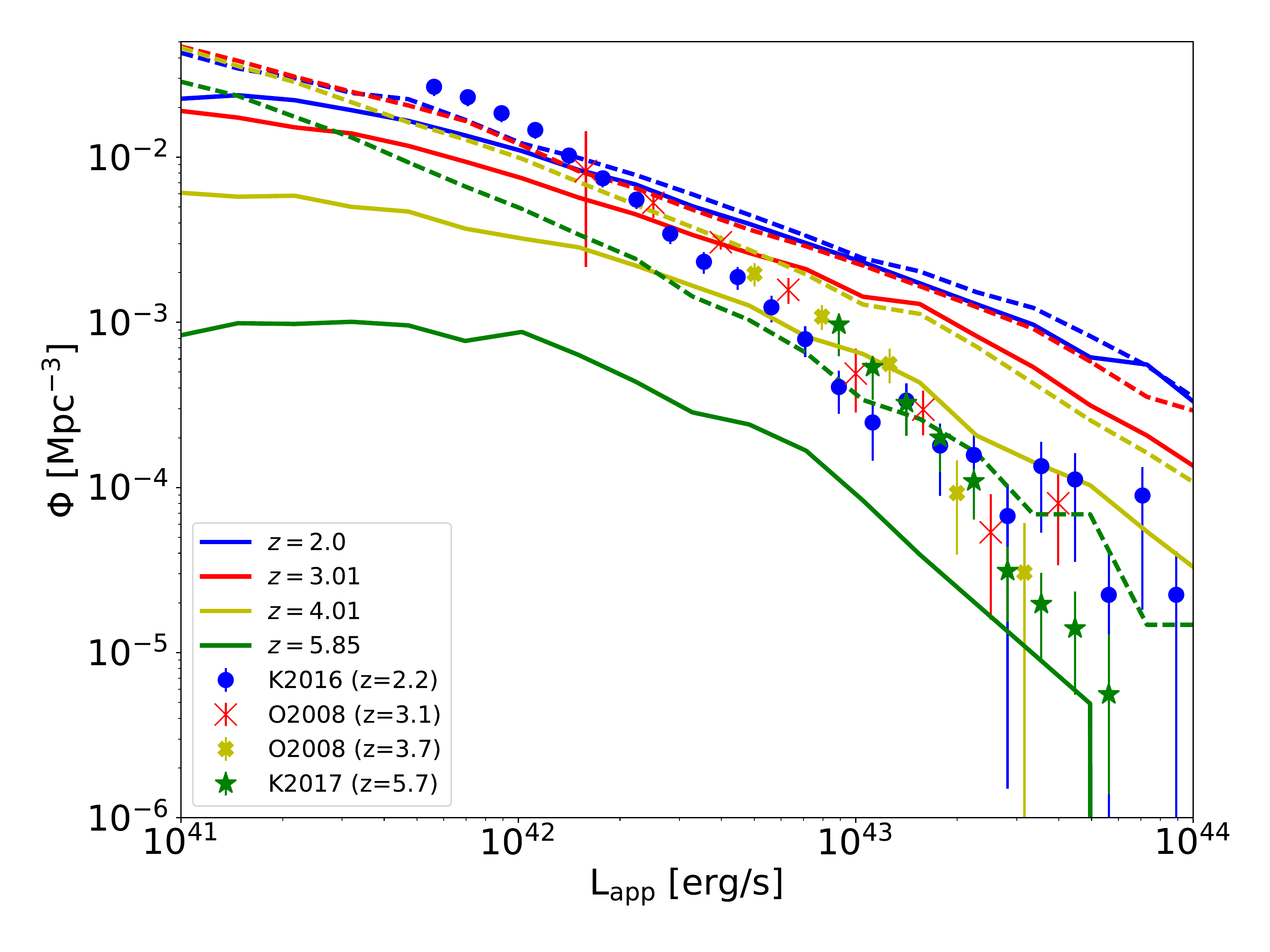}
      \caption{LAE luminosity function as obtained from an idealized observer after radiative transfer  using the fiducial emission model for different snapshots. Solid lines show the LF, dashed lines show the intrinsic LF. For comparison, observational data from \cite{Ouchi2008}, \cite{Konno2016}, and \cite{Konno2017} are shown.}
      \label{fig:apparentLF}
\end{figure}

In Fig. \ref{fig:apparentLF} we plot the LF after RT using our fiducial emitter model, corresponding to the LF an idealized observer would infer given the detection scheme outlined above for the high-resolution simulations. For comparison, we also plot the intrinsic luminosity (dashed) and observational data for the redshifts in question (symbols). While our $z=5.85$ LF is an order of magnitude too low, the synthetic LFs at low redshift have a high-luminosity tail not seen in observations. The latter can be attributed to the fact that we do not include dust in our simulations, which becomes increasingly important as progressing star formation enriches the ISM/CGM of galaxies at low redshifts. However, the underestimation of the LF at high redshifts is more difficult to understand. It can be expressed in terms of the average observed fraction, i.e., the ratio of the intrinsic Lyman-$\alpha$ luminosity to that inferred by an idealized observer $\epsilon$ averaged over all emitters. In our case it drops from about 80\% at $z=2.0$ to a few percent at $z=5.85$. We cannot attribute this tension to the cut in halo mass that we make. There are only roughly 1000 emitters at $z=5.85$ that are below the cut in halo mass but above the cut in SFR, and none of them has an intrinsic luminosity larger than 5 $\times$ 10$^{41}$.
Interestingly, ZZ11 also find their simulated LF to be too low compared to the observations although their intrinsic LF is much higher. This effect is related to the resolution of their simulation -- the mean observed fraction drops by a factor of approximately four if we employ a hydro resolution similar to the one used in ZZ11. 

In summary, we do not claim that our LAE sample well matches the observed LFs, because we oversimplify the physics to predict the apparent Lyman-$\alpha$ luminosity. In the following, therefore, we do not attempt to connect to the actual observed LAEs, but rather focus on the correlation between the large-scale environment and the apparent Lyman-$\alpha$ luminosity for our threshold LAE sample. 

\subsection{Correlations with large-scale structure}
In order to quantify the relation between the observed fraction of Lyman-$\alpha$ emission for galaxies and their large-scale environment, we follow ZZ11 and calculate the large-scale overdensity, $\delta$, its gradient with respect to the line-of-sight\footnote{we note that we rotate the coordinate system to have the line-of-sight aligned with the $z$ axis.} $z$, $\partial \delta/\partial z$, the velocity along the line of sight $v$, and its gradient along the line-of-sight, $\partial v /\partial z$. Since we are interested in the large-scale environment, we want to remove the small-scale structure in these datasets. To accomplish this, we follow ZZ11, and first retrieve the total density field on a fixed grid from the relevant snapshot. We  smooth the density field on a certain length scale (see below), and use this filtered density field to calculate the linear density gradient/velocity/velocity gradient fields. For this, we make use of the continuity equation and replace the time derivative involved by the growth factor. We stress that, identically to ZZ11 and BN13, we analyze the correlations with the large-scale environment using these large-scale smoothed fields. For the equations, see ZZ11, Eqs. 6-9.

\begin{figure*}
   \centering
   \includegraphics[width=\hsize]{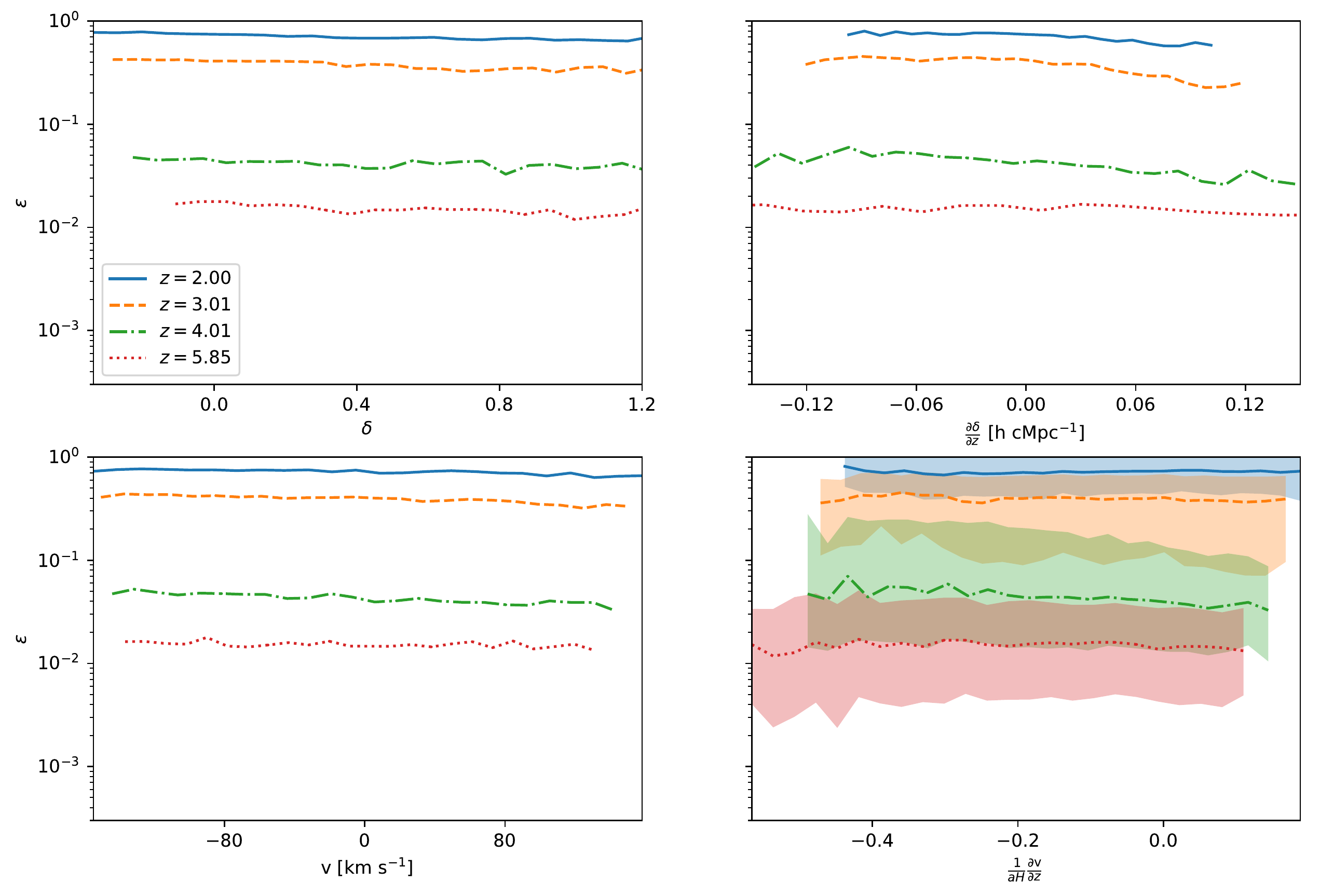}
      \caption{Correlation of median observed fraction $\epsilon$ of Lyman-$\alpha$ emission with large-scale overdensity $\delta$ (top left), velocity along the line of sight v (bottom left), and their respective derivatives with respect to the line of sight $\frac{\partial \delta}{\partial s}$/$\frac{\partial \mathrm{v}}{\partial s}$ (top right/bottom right) for different redshifts. In the lower-right panel, we show the quartile range for illustration. The scatter in the other panels is of similar magnitude.}
      \label{fig:corr_redshifts}
\end{figure*}

The filtering scales are estimated by comparing the full nonlinear power spectrum of each snapshot with the expected linear power spectrum, with the filter scale defined as the scale at which the two are different by more than 10\%. We choose scales of (3.6,2.7,2.1,1.3)  $h^{-1}$Mpc at redshift $z=$(2, 3.01, 4.01, 5.85). Figure \ref{fig:corr_redshifts} presents the resulting correlations at the four different redshifts, using a resolution of $\sim 1$ pkpc in the radiative transfer and our fiducial emitter model. 
On the $x$-axis, the different large-scale quantities are shown; Lines correspond to the binned median. In the case of the lower right panel, we also show the central two quartiles as shaded area to illustrate that not only is the evolution with the large-scale quantities very flat, but the scatter is very large. 
The offset between the lines for different redshifts is driven by the evolving mean observed fraction, which in turn is related to
the changing IGM neutral hydrogen density and expansion rate. We show the data only for one line of sight corresponding to the $x$-axis in the simulation volume. We note that the observed trend in the density gradient for $z = 3.01$ is not seen along other lines of sight and should therefore be taken with precaution. These results are broadly consistent with the results from BN13 using a smaller simulation volume. Nominally, the trends (e.g., with overdensity) are in the range of tens of percent, but as stated before much smaller than the scatter at fixed overdensity.

\begin{figure*}
   \centering
   \includegraphics[width=0.9\hsize]{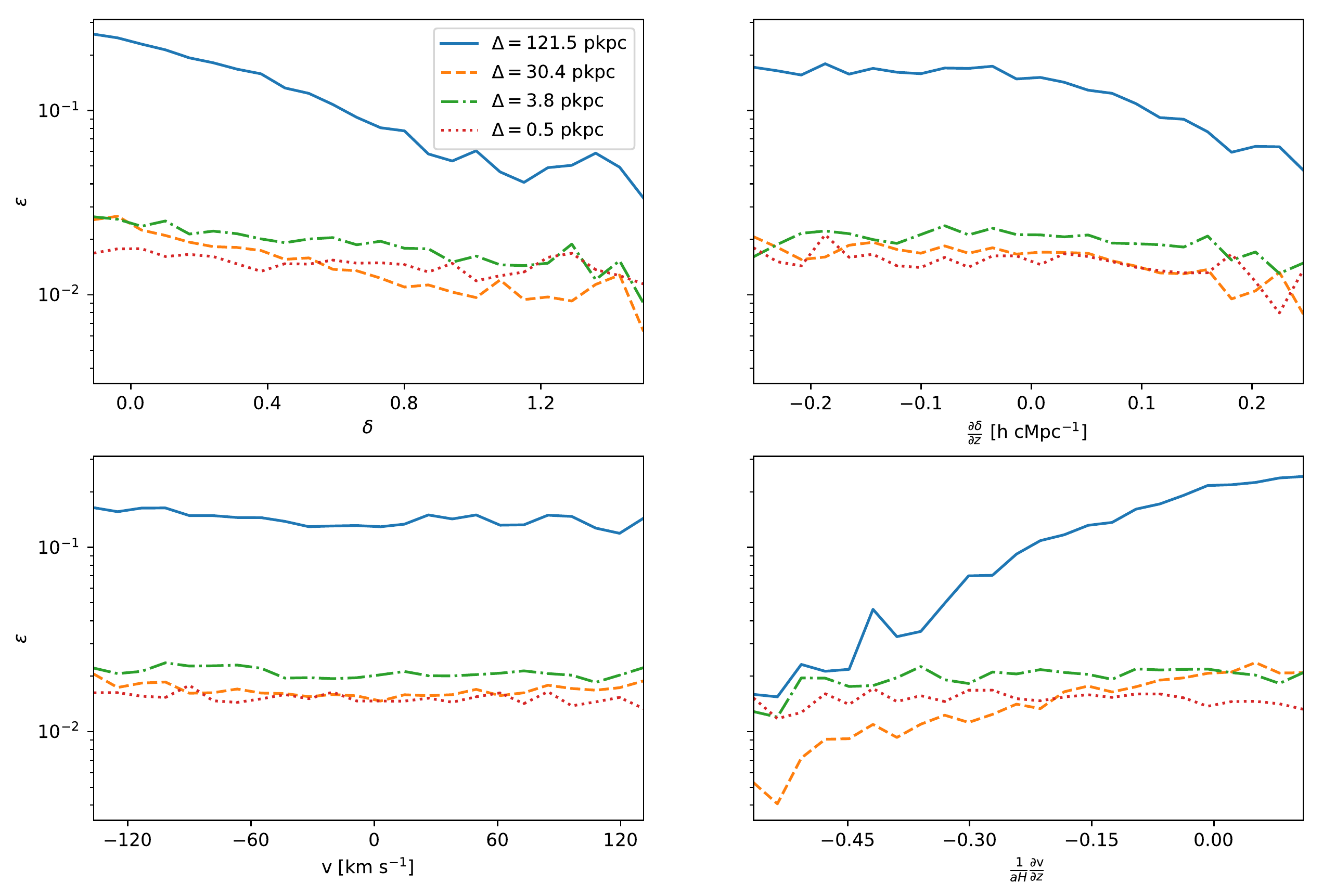}
      \caption{As in Fig. \ref{fig:corr_redshifts}, but for $z=5.85$ at four different spatial resolutions of the underlying RT simulation.}
      \label{fig:corr_resolution}
\end{figure*}

However, this picture changes dramatically if we degrade our spatial resolution. 
Figure \ref{fig:corr_resolution} shows the correlations 
for $z=5.85$ for four different resolutions. The highest and intermediate resolutions correspond to our fiducial choice and the resolution used in ZZ11, respectively. All correlations become stronger at lower resolutions. In particular, the line of sight velocity gradient has a dramatic impact on the observed fraction, modulating it by more than an order of magnitude. In the two low-resolution cases, these trends are much larger in amplitude than the associated scatter, consistent with the results of ZZ11. We note that even at lower redshifts (e.g. $z=3.01$), we find a significant trend if we artificially reduce our spatial resolution.

We note again that the average observed fraction changes nonmonotonically with resolution; it is highest for the lowest resolution case, with the medium resolution case having the smallest observed fraction. This effect is related to the size of the pixels of our synthetic detector. Reducing the size of pixels influences the mean observed fraction, but not the correlations with the large-scale structure as discussed above.
We therefore conclude that the main driver of the correlations seen in ZZ11 is the low spatial resolution of the underlying radiative transport simulation, leading to a nonseparation of galactic and intergalactic scales. At low resolution, galaxies and their surroundings are effectively smeared out on large scales, increasing the coupling between Lyman-$\alpha$ photons and the IGM. Although this effect depends on redshift -- higher redshift implies higher average cosmic density, in turn implying a stronger coupling -- we find it also at $z = 3.01$, albeit with a lower amplitude. This result of vanishing correlations of the observed fraction with the large-scale structure is in line with \cite{Behrens2013}, who found a similar result using the MareNostrum galaxy formation simulation at kpc resolution \citep{Ocvirk2008}.

Closer inspection reveals that the correlation found for lower resolution is related to the lower optical depth photons experience close to their emission spots, i.e., in the ISM. This is plausible, since the degraded resolution will primarily smooth out the small-scale ISM. The lower number of scatterings in the ISM yields, statistically,  a  smaller shift away from the line center, rendering the photons more susceptible to interactions with the IGM. This implies that the spectrum emerging from the ISM affects the strength of the correlations -- if we, for example, select only photons that escaped from the high-resolution simulations close to the line center (on the blue side), we can reproduce a correlation between observed fraction and  large-scale structure. This is in line with ZZ11 (see their Appendix D), who argued that a large shift from the line center can remove environmental effects due to lack of interaction with the IGM. However, their tests concerned only the intrinsic shift, that is, the line shift at emission, not the shift after processing by the ISM. 

Our argument that it is the processing in the ISM that drives the correlations is strengthened by the fact that the refinement structure of the large-scale filaments does not change over the range of the resolutions explored; as our refinement strategy is density-based, only regions of high density will reach the highest level of refinement. Indeed, we find that in filaments, we reach resolutions of $\sim$10-20 pkpc, independent of the maximum refinement level chosen \citep[we define filaments via their density, as in ][Table 2]{Haider2016}. 

We cannot strongly exclude that the small-scale structure of the ISM at scales of few parsec or even less will alter the line shift considerably, lacking simulations covering all the relevant scales from (sub)pc to comoving Mpc. However, we note that the trends become consistently weaker at all resolutions used in this work.

\begin{figure}
   \centering
   \includegraphics[width=\hsize]{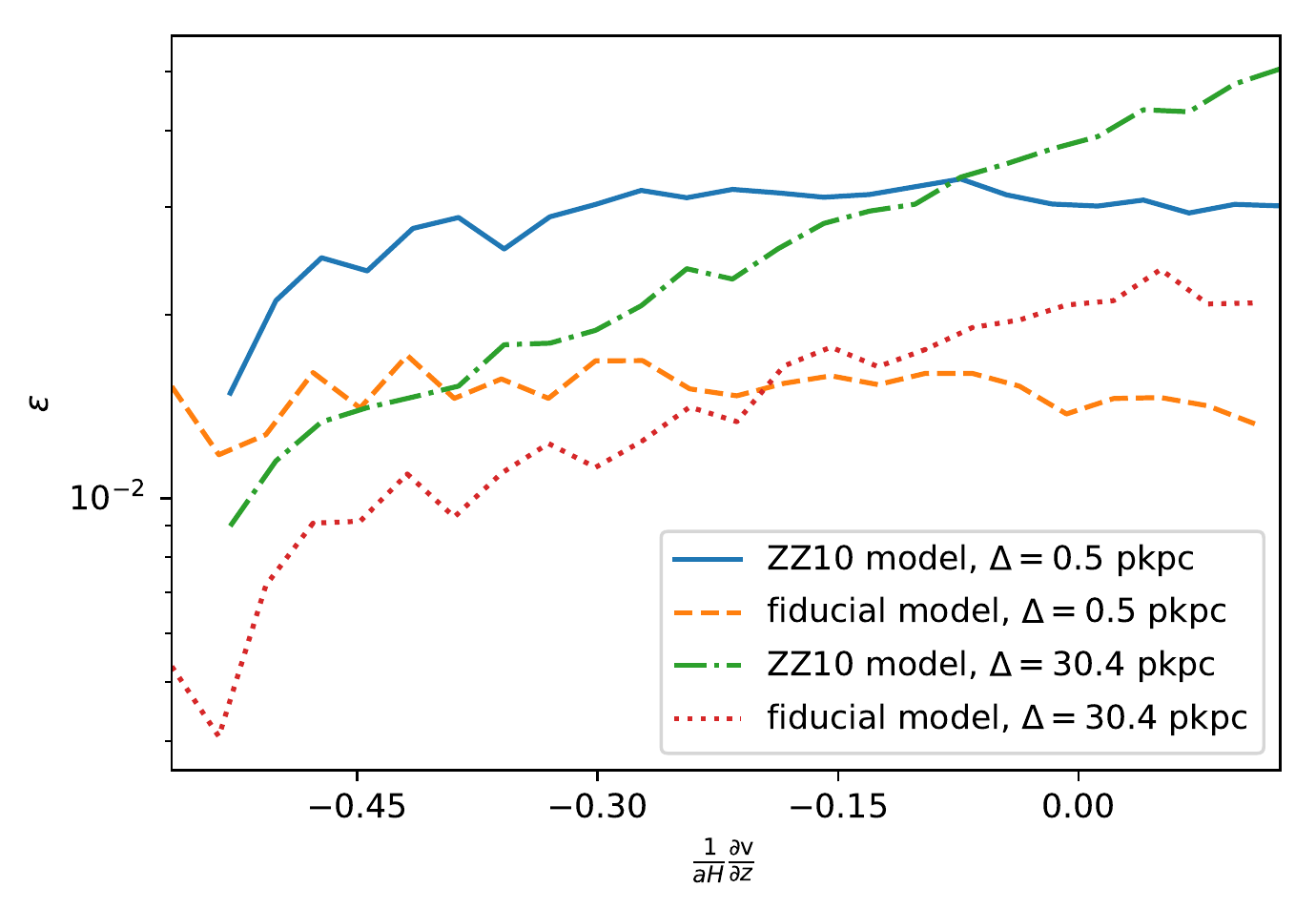}
      \caption{Correlation of observed fraction to large-scale velocity gradient for two different emitter models and resolutions.}
      \label{fig:corr_velgrad_zz}
\end{figure}

In addition, we notice that the employed emitter model can amplify the effect of low resolution on the correlations. When comparing with the ZZ11/BN13 emitter model, we again find no significant correlation at high spatial resolution. However, at low resolution, the correlations are more pronounced than for the fiducial emitter model, as shown in Fig. \ref{fig:corr_velgrad_zz} (again for $z=5.85$). We relate this to the fact we are statistically probing different environments due to the attribution of higher intrinsic luminosities to low-mass, less clustered halos in the ZZ11/BN13 emitter model.

Since we have rendered parts of the ISM transparent in our simulation,  to what extent this influences our results is a valid question. As we show in Appendix \ref{appendix:ism}, the mean observed fraction increases if we include the ISM as expected, but the correlations remain largely unchanged.

\subsection{Two-point correlation function}
\begin{figure*}
   \centering
   \includegraphics[width=\hsize]{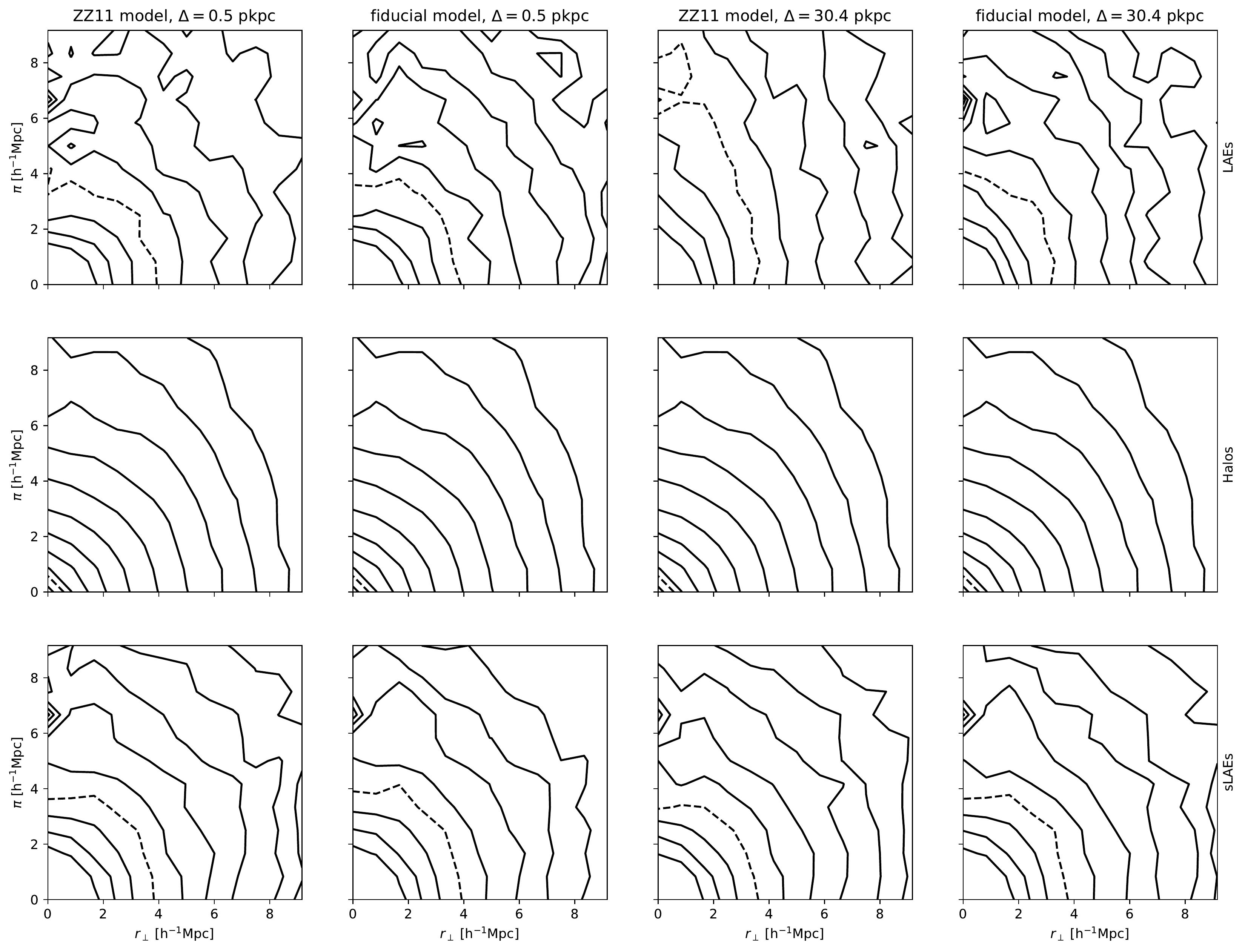}
      \caption{Two-dimensional 2PCF in real space for a number-density-limited sample of LAEs (first row) and two control samples (second/third row, see text for details). Columns show the results (from left to right) for a run with high resolution and the ZZ11 emitter model, high resolution and our fiducial emitter model, and the corresponding low-resolution runs with the ZZ11/fiducial emitter model. The dashed contour corresponds to $\xi=1$ with subsequent contours varying by a factor of $1.4$. We note the strong deformation in the third column.}
      \label{fig:2pcf}
\end{figure*}

The aforementioned two points - the influence of low spatial resolution and the emitter model - can readily be illustrated in terms of the 2PCF. 
The anisotropic clustering in 2PCF is well studied in the context of RSD. 
To isolate the selection effect due to the RT from RSD, we estimate the 2PCF in {\em real} space, following ZZ11. 
We estimate the two-dimensional 2PCF using the Landy-Szalay estimator \citep{Landy1993}, 
\begin{equation}
\xi({\bf r}) = \frac{DD(\Delta {\bf r})-2DR(\Delta {\bf r})+RR(\Delta {\bf r})}{RR(\Delta {\bf r})},
\end{equation}
where ${\bf r}=(r_{\perp}, \pi)$ is specified with the pair separation scale perpendicular to the line of sight, $r_{\perp}$, 
and that parallel to the line of sight, $\pi$, and $DD,\,DR,$ and $RR$ are number of counts in 
the data–data, data–random, and random–random pairs in a given bin, $[{\bf r}-\Delta{\bf r}/2,{\bf r}+\Delta{\bf r}/2]$.
We also quantify the anisotropy by the standard multipole moment defined by 
\begin{equation}
\xi_{\ell}(s) = \frac{2\ell+1}{2}\int^{1}_{-1}d\mu\,\xi(s,\mu)\mathcal{L}_{\ell}(\mu),
\end{equation}
where $s^{2}=r_{\perp}^{2}+\pi^{2}$, $\mu=\pi/s$, and $\mathcal{L}_{\ell}(\mu)$ is the Legendre polynomial at the $\ell$-th order. We utilize \texttt{halotools} \citep{hearin_high-precision_2016}\footnote{http://halotools.readthedocs.io/en/latest/} to compute the multipole 2PCF. We note that we adopt the slightly different definition on the multipole moment from BN13. 
In Fig. \ref{fig:2pcf}, we show the 2PCF for the simulation at $z=5.85$, varying both ingredients (columns) - the emitter model and the spatial resolution.
The three rows show the 2PCF of halos, LAEs, and a shuffled LAE sample. Similar to previous work, halo and LAE samples are acquired by enforcing a certain total number density, which we set to 10$^{-2}$h$^{3}$ Mpc$^{-3}$ here. The result is robust for varying number densities and thus a high density has been chosen to reduce noise. The shuffled LAE sample is generated by shuffling the inferred luminosity among galaxies of similar mass (by sorting the halos into mass bins of width $0.03$ dex) in order to erase any effect introduced by the radiative transport, and serves as another control sample.

\begin{figure*}
   \centering
   \includegraphics[width=1.0\hsize]{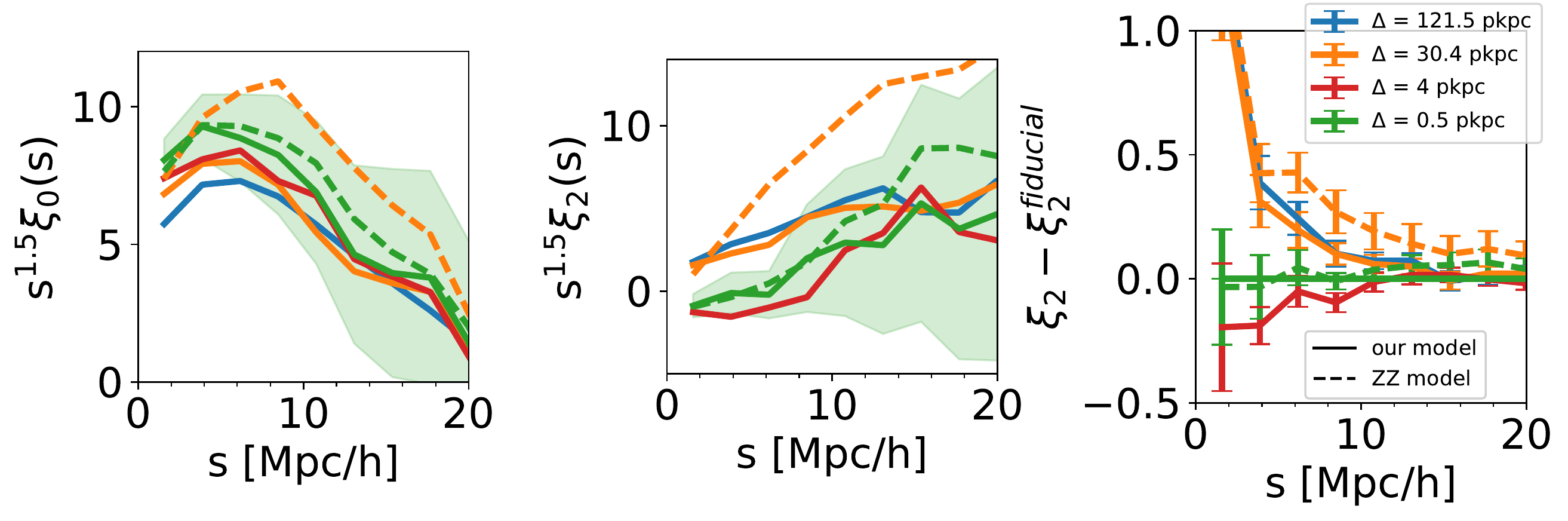}
      \caption{Real-space monopole (left panel), quadrupole (middle), and the difference in the quadrupole between our high-resolution run employing the fiducial emitter model and the other samples (right), for the distribution of LAEs in the number density samples for $z=5.85$. Simulations with different spatial resolutions (high: green, orange: intermediate, blue: low) and for both the fiducial (solid) and the ZZ11 emitter model (dashed) are shown. Errors were estimated using a Jackknife approach and are only shown for the fiducial high-resolution dataset for illustration in the left and center panels. For the right panel, the differential error is shown, reducing the sampling variance.}
      \label{fig:quadrupole}
\end{figure*}

As expected from the discussion above, we find a significant deformation of the shape of the 2PCF only at low resolution. The strength of the deformation further increases in the case of the ZZ11 emitter model. However, we stress that at our fiducial high resolution, we do not find a deformation of the 2PCF even for this emitter model. We therefore conclude that the critical ingredient here indeed is the spatial resolution of the underlying simulation.
As a more quantitative probe of the resulting deformation, we show the monopole (left panel) quadrupole (middle panel) moment, and the difference between the quadrupole in our high-resolution simulation with the fiducial emitter model and the other simulation runs (right panel) in Fig. \ref{fig:quadrupole}. Different colors represent different spatial resolutions. Solid lines show the results for the fiducial emitter model, while dashed lines are obtained using the ZZ11 emitter model. The shaded regions in the left and middle panel illustrate the statistical errors (including sampling variance taken from the covariance matrix estimated by 64 subsamples). We note that the jackknife errors may tend to overestimate the true ones \citep{Norberg2009,Shirasaki2016}. Nevertheless this does not affect our main conclusion because we are mainly interested in the relative difference in Fig. \ref{fig:quadrupole} (right panel), for which we show the relative error, greatly reducing the sample variance. At $s<10$ Mpc/h, the lower-resolution runs (orange, blue) show a significant increase in the quadrupole, as expected from the 2PCF. For the ZZ emitter model, we find an even stronger increase, again consistent with the expectations.

\section{Conclusions}
\label{sec:conclusions}
Using data from the public release of the Illustris suite of cosmological simulation, we have run Lyman-$\alpha$ radiative transfer simulations as a post-processing step at redshift of $2<z<6$ to understand if and to what extent the complex radiative transfer introduces an additional, anisotropic selection bias into measurements of the large-scale clustering of galaxies relying on the Lyman-$\alpha$ emission line.

At sufficiently high spatial resolution down to $\mathcal{O}(1)$ physical kpc, we find only marginal correlations between the large-scale density and velocity fields with 
the observed fraction of Lyman-$\alpha$ emission, and consequently smaller impact on the resulting 2PCF than what is previously claimed by ZZ11. This result is in line with our previous finding in a smaller simulation box presented in BN13.

We were able to reproduce the ZZ11 result by reducing the spatial resolution of our simulation to a value close to the simulations used by ZZ11. We therefore conclude that the discrepancy is due to incomplete scale separation, effectively decreasing the density of the ISM, coupling the photons more strongly to the large-scale environment. This fact simply implies that the impact of RT is mainly driven by the fully nonlinear scale. We also find that this resolution effect can be amplified by the choice of an intrinsic Ly$\alpha$ emitter model with many low-luminosity, low-mass emitters. 

A caveat of this analysis is related to our finding that the detailed spectra emerging from the ISM have a strong influence on the coupling between photons and the large-scale structure in line with \cite{Laursen2011}; if the spectra are close to the line center, this could reintroduce the correlation between observed fraction and large-scale velocity gradient. Future work will be able to quantitatively assess this relation. Ultimately, there is a need for very high-resolution hydro/RT simulations (that is, sub-pc resolution) to obtain a realistic picture of RT through the small-scale structure of the ISM \citep[see][]{Gronke2016b}. Additionally, the influence of dust might be important here since dust tends to preferentially dampen the spectrum far away from the line center \citep{Laursen2009}. Due to the observed relations between escape fraction (that is, the fraction of Lyman-$\alpha$ able to leave the ISM without being absorbed by dust) and mass, the effect of dust on these large-scale correlations is not obvious \citep{Garel2012}.

Although our primary focus in this paper was the impact of RT on the anisotropic selection bias in the LAE clustering, our RT code can be in principle applied to other galaxy formation simulations, and our simulation sets shall be useful for other studies as well. For example, it would be challenging but desirable to apply our RT code to the improved IllustrisTNG simulation \citep{Springel2017}. IllustrisTNG has a larger box size but the same mass resolution, allowing us to extend our statistics to larger scales for the BAO and RSD measurements. Even our current simulation sets may shed light on the relation between LAEs and their host dark matter halos \citep[e.g.,][]{Tilvi2009,Nagamine2010,Ouchi2010,Ouchi2017,Kusakabe2017}, Ly$\alpha$ intensity mapping \citep[e.g.,][]{Pullen2013,Comaschi2016,Fonseca2017}, Ly$\alpha$ forest (e.g., Kakiichi \& Dijikstra in prep.) or the cosmic reionization \citep[e.g.,][]{Yajima2017}. We hope to address these in the near future and note that our datasets are available upon request.

\begin{acknowledgements}
We thank Koki Kakiichi and Eiichiro Komatsu for numerous fruitful discussions.
We also thank D. Nelson for help in using the Illustris data, and J.F. Engels as well as H. Braun for useful discussions. 
This work was supported in part by JSPS KAKENHI Grant Number JP15H05896.
We acknowledge use of the Python programming language \citep{VanRossum1991}, and use of the Numpy \citep{VanDerWalt2011}, IPython \citep{Perez2007}, and Matplotlib \citep{Hunter2007} packages. This research made use of Astropy, a community-developed core Python package for Astronomy \citep{astropy}, and halotools \citep{hearin_high-precision_2016} for the computation of the TPCFs.
\end{acknowledgements}


\bibliographystyle{aa}

\bibliography{LyA_Illustris_I}

\begin{appendix}
\section{Details of the RT calculations}
\subsection{Implementation of the Hubble flow}
In principle, the Hubble flow in-between two scatterings of a Lyman-$\alpha$ photon will simply redshift the photon by
\begin{equation}
\Delta x = - \frac{H(z)d}{v_{th}}
,\end{equation}
with the usual definition of a dimensionless frequency, $x = \frac{\nu_0-\nu}{\Delta \nu_D}$, $\nu_0$ the line center frequency of Lyman-$\alpha$, $\nu$ the photons frequency, $\Delta \nu_D$ the Doppler frequency of the gas, $v_{th}$ the thermal velocity of the gas, and $H(z)$ the Hubble factor at redshift $z$. As \cite{Semelin2007} point out, this is a good approximation at low redshift (<10) and for relatively small cosmological distances (< 100 Mpc). However, since the Voigt profile determining the optical depth a photon sees is a very steep function close to the line center, care must be taken to ensure that the shift of the frequency is evaluated with sufficient spatial sampling. By default, our code transports photons AMR cell by AMR cell, that is, through one cell per step. If the cells are small enough (e.g., several kpc), it is sufficient to evaluate the Voigt profile at the frequency of the photon once per cell; However, in large cells, this might result in a large error on the optical depth in this cell, for example if the photon in question hits the line center frequency within the cell in question. To avoid this problem, we set a fixed maximum step size of 10$^{-4}$ corresponding to about 10 kpc, that is, the evolving cross section within each cell is evaluated at a spatial resolution of 10 kpc.
\subsection{Influence of the acceleration scheme}
To speed up the RT calculations, it is common to employ a so-called acceleration scheme to avoid core scatterings, that is, scatterings from the line center into the line center, because in the regime of high optical depth, these do not affect the outcome of the RT calculation, but increase the computational cost. While there are refined schemes like the one proposed by \cite{Laursen2010b}, frequently a hard-coded cutoff is used to avoid scatterings with $|x|<3$, with $x$ the frequency of the photon in units of the Doppler frequency. We tested both the Laursen scheme and the hard-coded cutoff and found no significant difference. At low resolution, the constant cut-off tends to overestimate the observed fraction by about 10\%. However, this has no significant effect on the correlations discussed in this paper, and we therefore present only the results for the hard-coded cutoff.
\subsection{Convergence}
\begin{figure*}
   \centering
   \includegraphics[width=0.9\hsize]{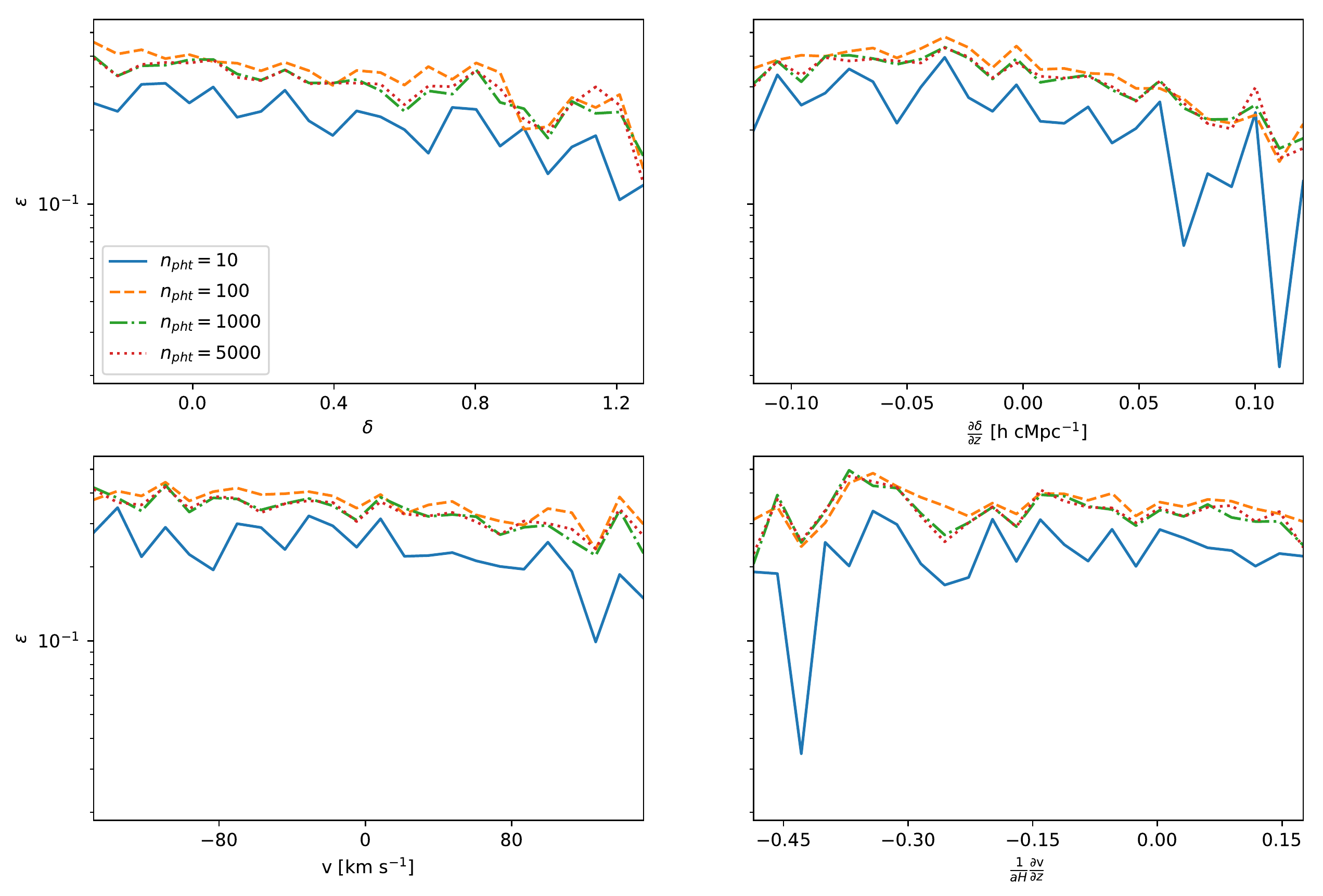}
      \caption{As in Fig. \ref{fig:corr_redshifts}, but for $z=3.01$ for different values of $n_{pht}$.}
      \label{fig:convergence}
\end{figure*}

Similar to ZZ11/BN13, we specified a minimum number of tracer photons launched from every emitter. We checked explicitly if the correlations and observed fraction analyzed here are converged in our simulations: in Fig. \ref{fig:convergence}, we show the correlations at $z=3.01$ and vary the minimum number of photons per halo, $n_{pht}$, from 10 to 5000 for a random subsample of 5000 emitters. The correlations (and the observed fractions) are sufficiently converged at $n_{pht}=100$, thus all plots shown in this work are based on simulations using this number. We also ran a number of full simulations with $n_{pht}=100$ and do not find to differ significantly within the scope of this paper; due to the size of the output data, we stick to the smaller value, since the data is easier to handle. We stress, however, that the spectra of individual emitters, for example, are very noisy with a very low number of photons per halo.

\subsection{Influence of $\tau_{max}$ and d$_{cut}$}
As discussed above, we follow photons only for a fraction d$_{cut}$ of the box size before we assume the radiative transfer to be complete. The motivation here is the fact that the Hubble flow quickly transports photons out of resonance, reducing the optical depth drastically once the photons have travelled several Mpc. To check that this argument is correct, we reran the $z=3.01$ case with a different value for d$_{cut}$. We compared the distribution of observed fractions for our fiducial value of d$_{cut}=0.3$ and a larger value of 0.9, and found no significant differences.

Additionally, we also checked the influence of the maximum optical depth $\tau_{max}$ that is considered in the peeling Off scheme before discarding a photon. By default, we set $\tau_{max}=20$. Setting $\tau_{max}$ to a higher value of 30 instead a distribution of observed fractions almost identical to the fiducial one.

\section{Influence of the inclusion/exclusion of the ISM}
\label{appendix:ism}
\begin{figure*}
   \centering
   \includegraphics[width=0.9\hsize]{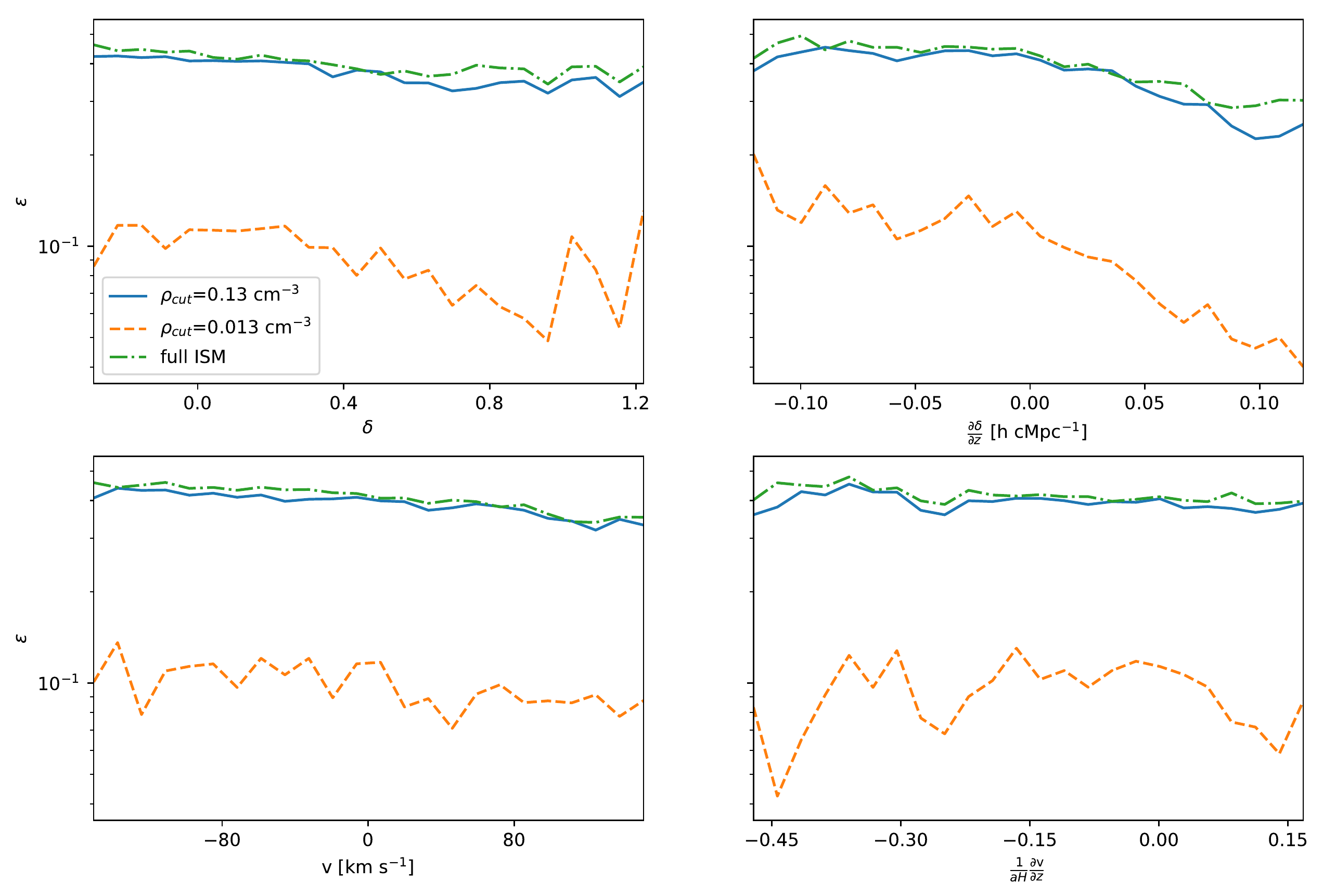}
      \caption{As in Fig. \ref{fig:corr_redshifts}, but for $z=3.01$ for three different values of $\rho_{\mathrm{cut}}$.}
      \label{fig:ISM}
\end{figure*}

As stated above, in this work we removed the ISM, defined by the density threshold used in Illustris for star-forming gas, from the simulation prior to running the radiative transfer. However, for comparison we also ran simulations including the full ISM structure, and simulations with more of the ISM removed. We find that while the mean observed fraction is higher in this case. The higher mean observed fraction has an intuitive explanation; scatterings in the ISM tend to shift photons away from resonance, reducing their optical depth in the immediate surroundings. However, the correlations with the large-scale structure do not change significantly. As an example, we show the results for the correlations at $z=3.01$ for three different threshold of $\rho_{\mathrm{cut}}$ in Fig. \ref{fig:ISM}.

\end{appendix} 

\end{document}